\documentstyle[twoside,fleqn,espcrc2,epsf]{article}

\newcommand{\AmS}{{\protect\the\textfont2
  A\kern-.1667em\lower.5ex\hbox{M}\kern-.125emS}}

\title{The spectral dimension of non-generic branched polymers}

\author{John F. Wheater\address{Theoretical Physics,\\
        1 Keble Road, \\
        Oxford OX1 3NP, United Kingdom}%
        \thanks{Speaker at the conference}
        and
        Joao Correia
\address{Theoretical Physics,\\
       1 Keble Road, \\
       Oxford OX1 3NP, United Kingdom}
}

\begin{document}

\begin{abstract}
We show that the spectral dimension on non-generic branched polymers
with susceptibility exponent $\gamma>0$ is given by
$d_s=2/(1+\gamma)$. For those models with $\gamma<0$ we find that
$d_s=2$.
\end{abstract}

\maketitle

\section{INTRODUCTION}
The manifolds in the ensemble of two-dimensional 
quantum gravity have a rich structure which can be characterized
in a number of different ways. In particular there are
several different quantities which in the case of smooth regular manifolds
take the same value and correspond to the usual notion of dimension.
When computed for manifolds which are far from regular these quantities,
which in fact probe different aspects of the geometry,
 can yield different values.

The Hausdorff dimension $d_h$ is defined
         through the volume, $dV$ of an annulus
of geodesic thickness $dR$ at geodesic distance $R$ from a point by 
\begin{equation} dV \sim R^{d_h-1}\,dR\end{equation}
On the other hand the spectral dimension $d_s$ describes the properties of
a random walk, generated by the diffusion equation, on the manifold. A walker
sets off from a point P; the probability that after making $t$ steps
(a step is defined as a move from one lattice point to a neighbouring one
in discretized quantum gravity) he has returned to P is
\begin{equation}P(t)\sim\frac{1}{t^{d_s/2}}\end{equation}
provided that $N^\Delta \gg t\gg 1$ to avoid finite size and discretization 
effects respectively. For smooth regular two dimensional manifolds
$d_h=d_s=2$. The spectral dimension has been determined for various 
two-dimensional quantum gravity systems by Monte Carlo simulation, scaling 
relations, and exact calculation. In \cite{generic} 
it was shown that for the generic 
branched polymer ensemble $d_s=4/3$ whereas the corresponding Hausdorff 
dimension is 2 \cite{BP}. Using the same method we have now 
extended the calculation to the non-generic branched polymers \cite{us}.

\section{CALCULATION}
The branched polymer ensembles have a grand canonical partition function
$Z(\mu)$  given by \cite{BP}
\begin{equation}Z(\mu)=
e^{-\mu}(1+\sum_{n=1}^\infty\alpha_n Z(\mu)^n)\end{equation}
which is non-analytic as $\mu\downarrow\mu_{cr}$ with behaviour
\begin{equation}Z(\mu)\simeq Const  - (\mu-\mu_{cr})^{1-\gamma}\end{equation}
For generic branched polymers (a finite number of $\alpha_n$ non-zero, and all
positive) we get $\gamma=1/2$; the multi-critical BPs (at least $k$ $\alpha$s
non-zero and specially tuned) have $\gamma=1-1/(k+1)$ with $k=2,3,\ldots$
and the continuous critical BPs have any $\gamma < 1$ at the expense of
having an infinite number of the $\alpha$s non-zero.

\begin{figure}[h]
\vspace{9pt}
{\epsfxsize=7cm \epsfbox{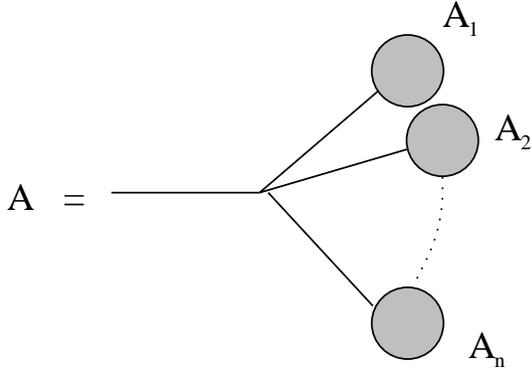}}
\caption{The constituents ${\rm A}_1,\ldots,{\rm A}_n$  of a branched polymer
$\rm A$.}
\end{figure}

Our calculation of $ d_s$ for the non-generic BPs \cite{us} 
follows the method of \cite{generic}; the main generalization is that the
 generating function for the return
probabilities on {\em any} branched polymer A can always be expressed in terms
of those for its constituents ${\rm A}_1,{\rm A}_2,\ldots$
(see fig.1). Letting 
\begin{equation}P_{{\rm A}}(y)=\sum_t y^t P(t)=\frac{1}{1-y}\frac{1}{h_{{\rm A}}(y)}\end{equation}
we find that 
\begin{equation}h_{{\rm A}}(y)=\frac{1+\sum_{i=1}^n h_{{\rm A}_i}(y)}
{1+(1-y)\sum_{i=1}^n h_{{\rm A}_i}(y)}\end{equation}
The main result that we  prove is that defining
\begin{equation}H_{{\rm A}}^{(n_1,n_2,\ldots,n_p)}=
\prod_{i=1}^p\frac{d^{n_i}}{dy^{n_i}}h_{{\rm A}}(y)\vert_{y=1}\end{equation}
then 
\begin{equation}H^{(n_1,n_2,\ldots,n_p)}(z)=
\sum_{{\rm A}} z^{N_{{\rm A}}} w_{{\rm A}}H_{{\rm A}}^{(n_1,n_2,\ldots,n_p)}
\end{equation}
has the asymptotic behaviour
\begin{equation}H^{(n_1,n_2,\ldots,n_p)}(z)\sim(z_{cr}-z)^{1-\gamma-p-(1+\gamma)\sum_{i=1}^pn_i}\end{equation}
(where $z=e^{-\mu}$, the sum over A runs over all BPs, $N_{{\rm A}}$
is the number of links in A, and $w_{{\rm A}}$ the weight for A in the 
graphical expansion of (3)). This is done by induction on $n$ and $p$ and using
(6) to generate the derivatives.

Using this result we can compute the asymptotic expansion of the
grand canonical generating function
\begin{equation} {\cal P}(z,y)=\sum_{{\rm A}} z^{N_{{\rm A}}}
 w_{{\rm A}}P_{{\rm A}}(y)\end{equation}
Taking the n'th derivative of (10) and setting $y=1$ yields a sum of 
terms which are precisely the $ H^{(n_1,n_2,\ldots,n_p)}(z)$. Thus we are 
able to show that 
(supposing for a moment that the Taylor series is convergent about $y=1$,
 and dropping a trivial pole term)
\begin{equation}{\cal P}(z,y)=\left(1-z/z_{cr}\right)^\beta\Phi\left(
\frac{1-y}{\left(1-z/z_{cr}\right)^\Delta}\right)\end{equation}
with $\beta=1-2\gamma$ and $\Delta=1+\gamma$.  In fact the Taylor series
is not convergent, nor is it Borel summable. However it is possible to
prove \cite{generic} that it is an asymptotic series with an
 integral representation
which has no unphysical singularities in the region of interest. The absence
of Borel summability means that this integral representation is not unique; 
however it can only differ from the true result by terms involving essential
singularities which vanish at $y=1$. Since $\cal P$ has ordinary non-analytic
behaviour at $y=1$ these possible essential singularities can have no 
effect on the physics.

\section{RESULTS}

It was shown in \cite{us} that the critical behaviour of $\cal P$ is
related to the spectral dimension through
\begin{equation} \beta=1-\gamma+\Delta\left(\frac{d_s}{2}-1\right)
\end{equation}
When $\gamma>0$ we find that $d_s=2/(1+\gamma)$. This is of course 
consistent with the previously known result for the generic case. It also
shows that the spectral dimension is not an independent critical exponent
for these models -- it is completely determined by $\gamma$. For the models
with $\gamma <0$ (for which $\Delta=1$ and $\beta=1-\gamma$) we find that $d_s=2$ always. This is the same as the value
now established for the quantum gravity phase \cite{rolf}  but this is a
coincidence. The negative $\gamma$ BPs are rather pathological objects being
very bushy (this is the effect of the dominance of the
 high order branching in (3)) and there is no sense in which the walker is
really ``diffusing''; in reality he is just going backwards and forwards
along the short branches of the bush. One cannot deduce from the coincidence
of $d_s$ that negative $\gamma$ BPs are the same as the quantum gravity phase!


\begin{thebibliography}{9}
\bibitem{generic} T.Jonsson, J.F.Wheater, Nucl. Phys. B 515 (1998) 549.


\bibitem{us} J.D.Correia, J.F.Wheater,  Phys. Lett. B 422 (1998) 76.

\bibitem{BP} J.Ambj\o rn, B.Durhuus, T.Jonsson, Phys. Lett. B 244 (1990) 403.
\bibitem{rolf}  J. Ambjorn, D. Boulatov, J.L. Nielsen, J. Rolf, Y. Watabiki,
J.High Energy Phys. 02 (1998) 010.
\end{thebibliography}
\end{document}